\documentclass[%
 reprint,
 amsmath,amssymb,
 aps,
]{revtex4-1}

\usepackage{graphicx}
\usepackage{dcolumn}
\usepackage{bm}

\begin{document}

\title{D meson production in Pb-Pb collisions with the ALICE detector}%

\author{Riccardo Russo}

 \email{rrusso@to.infn.it}
\affiliation{Universita' degli studi di Torino \\ Istituto Nazionale di Fisica Nucleare, Sezione di Torino}
\collaboration{ALICE Collaboration}

\date{\today}

\begin{abstract}

Open heavy-flavour hadrons
are a powerful tool to investigate the properties of the high-density medium
created in heavy-ion collisions at high energies
 as they come from the hadronization of heavy quarks. The latter are created 
in the early stage of the interaction and experience the whole collision history.
Heavy quarks in-medium energy loss can be investigated by comparing the heavy flavour production cross sections in p-p and nucleus-nucleus
collisions. In addition, initial spatial anisotropy of the fireball is converted into momentum
anysotropy of final state particles, in particular the elliptic flow.
D mesons  are identified from their  hadronic decays 
which can be  reconstructed in the central rapidity region using 
the tracking and PID detectors.
We report on the measurements of  
${\rm D^+}$, ${\rm D^0}$, ${\rm D^{*+}}$ and ${\rm D_s^+}$ production as a function of transverse momentum in pp collisions at  $\sqrt s$ = 7 TeV and Pb-Pb collisions at $\sqrt {s_{\rm NN}}$ = 2.76 TeV which
 allow one to calculate the nuclear
modification factor expected to be sensitive to the in-medium
energy loss of charm quarks, and on the elliptic flow in central and semiperipheral collisions.
\end{abstract}

\maketitle


\section{\label{sec:level1} Open Heavy Flavours as a probe of QGP formation}
Different QCD-based models, such as lattice QCD~\cite{lattice}, predict that at extremely high density or temperature the nuclear matter undergoes a phase 
transition to a state called Quark Gluon Plasma (QGP) in which quarks and gluons are no more bound inside hadrons, but behave as free particles over large volumes. The QGP is characterized by a larger number of degrees of freedom than ordinary nuclear matter, therefore larger energy and entropy densities are expected.\\
Such conditions can be experimentally created in heavy-ion collisions, where the large number of participant nucleons results in the creation
 of a fireball characterized by large volume (several fm/c) and number of constituents. The fireball expands 
reaching  the phase boundary between QGP and  hadron gas. The properties of the QGP  can be investigated through the study of
many different observables, which can be reconstructed in the experimental apparatus, in our case ALICE. \\
In particular, D mesons can give information about the evolution of charm quarks in the medium. \\
Charm quarks are produced in  high-$Q^2$ scattering processes. Given their short production timescale ($\sim 1/Q^2$), they are expected to experience the full evolution 
of the created medium before hadronizing.\\
In particular, it is expected that
partons, while traversing the medium, lose part of their energy via elastic
collisions with the other constituents and gluon radiation~\cite{enloss}. This
in-medium energy loss can be studied by comparing transverse momentum ($p_{\rm T}$) spectra of D mesons 
in AA and pp collisions, where no medium is produced, by means of the nuclear modification factor R$_{\rm AA}$ 
\begin{center}
\scriptsize
$R_{AA}(p_{\rm T})=\frac{1}{N_{\rm coll}}\frac{dN_{\rm AA}/dp_{\rm T}}{dN_{\rm pp}/dp_{\rm T}}$
\end{center}
where N$_{\rm coll}$ is the average number of binary collisions that occur in a single nucleus-nucleus collision, N$_{\rm AA}$ and N$_{\rm pp}$ are the 
measured D meson yields in AA and pp collisions respectively.
R$_{\rm AA}$=1 is expected if heavy quarks production is not affected by the hot medium. Energy loss causes
R$_{\rm AA}<$1 at intermediate and high $p_{\rm T}$.\\
  Gluon radiation 
is a QCD effect and is expected to depend on the parton color charge: gluons are  expected to lose more energy than quarks, so D mesons should
show a lower suppression than light charged hadrons, mainly coming from gluon fragmentation.  A
suppression of gluon radiation at small angles relative to the parton
momentum is also expected. The angle of this ``dead cone'' is expected to increase with the parton mass~\cite{deadcone}. As a consequence,
  hadrons originating from b quarks are expected to be less suppressed
 than those coming from c quarks which in turn are less suppressed than light hadrons. In addition, the relative yield of $D_s^+$
 with respect
to that of non-strange D mesons is expected
to be
enhanced
in Pb-Pb collisions
at
low-intermediate $p_{\rm T}$
if charm quarks
hadronize via
recombination
in the medium, due to  strangeness enhancement in AA collisions~\cite{ds}~\cite{ds1}.\\
The interpretation of the $R_{\rm AA}$ measurements  has also to take into account the presence of initial state effects in AA collisions, such as nuclear modification of the Parton Distribution
Functions 
(shadowing, anti-shadowing, EMC effect) and gluon saturation which would affect the production of
c-quarks in heavy-ion collisions. These effects can be also observed in pA collisions, where no medium is created. The analysis on the 2013
p-Pb data sample at the LHC is currently ongoing.
\\
Another observable  that will be discussed in these
proceedings is the elliptic flow of D mesons. In non-central collisions, the spatial anisotropy of the  overlap region of the colliding nuclei is converted into momentum
anisotropy of final state particles due to interactions
among the medium constituents~\cite{v2}. The anisotropy is quantified by the coefficients of the Fourier
expansion of the distribution of the final state particles azimuthal angles
relative to the reaction plane, which is defined by the
impact parameter  of the collision and the beam direction \\
\begin{center}
\scriptsize
 $\frac{dN}{d(\varphi-\Psi_{\rm RP})} \propto 1 + 2v_{1}\cos(\varphi-\Psi_{RP})+ 2v_{2}\cos(2(\varphi-\Psi_{\rm RP})) + ...$
\end{center}
In particular, $v_{2}$ is called elliptic flow and it reflects the initial almond-shaped
geometry in non-central collisions.\\
The study of D meson $v_2$ is a tool to understand how c quarks interact with the other constituents and participate to the collective expansion of the fireball.
 In addition, a positive $v_{2}$ 
is expected at high $p_{\rm T}$ due to the path-length dependence of energy loss, as c quarks emitted  perpendicular to the reaction
plane trasverse a longer distance in the medium.
\section{\label{sec:level1} Data sample and analysis}
All the details on the ALICE detector can be found in Ref. ~\cite{jinst}. \\
The results presented here are obtained from the 2011 Pb-Pb data sample at $\sqrt{s_{\rm NN}}$ = 2.76 TeV.
Data have been collected with a minimum bias trigger, based on the VZERO detector (array of scintillators covering the
full azimuth at -3.7$<\eta<$-1.7 and 2.8$<\eta<$5.1); to enhance 
the number of events in the desired centrality classes, an online selection based on the VZERO signal amplitude has been used. Only events
with a vertex found within 10 cm from the centre of the detector along the beam axis were used.
\\

The analysed sample consists of 16$\times 10^6$ events
in the centrality class 0-7.5\% corresponding to an integrated luminosity of
28 $\mu$b$^{-1}$. For the more peripheral samples, 9.5$\times 10^6$ events in 30-50\%, and
7.1$\times 10^6$ in 15-30\%, corresponding to an integrated luminosity of 6 $\mu$b$^{-1}$, were
analyzed.


\subsection{\label{sec:level2} Nuclear modification factor}
The analysis is based on the reconstruction of D mesons in the following  hadronic decay channels $D^{0}\rightarrow K^{-}\pi^{+}$,
 $D^{+}\rightarrow K^{-}\pi^{+}\pi^{+}$,
 $D^{*+}\rightarrow D^{0}\pi^{+}$ and
 $D_{s}^{+}\rightarrow \phi\pi^{+}\rightarrow K^{+}K^{-}\pi^{+}$  in the ALICE central barrel ($|\eta|<$0.9). The Inner Tracking System (ITS) and 
Time Projection Chamber (TPC) provide 
excellent tracking performance, with impact parameter resolution of few tens of $\mu$m for  $p_{\rm T}>$1 GeV/c 
and transverse momentum resolution of better than 2\% up to 10 GeV/c for Pb-Pb collisions. \\
D meson candidates are formed by combining pairs and triplets of tracks within each event.

Topological cuts are applied to reduce the large
combinatorial background, requiring in particular a significant separation
between the primary and secondary vertices (c$\tau$ $\sim$ 100-300 depending on the D meson species) and small
 pointing angle (angle 
between the reconstructed momentum and the D meson flight line).


 A further background rejection is obtained applying a PID 
selection  on the decay tracks by combining information from the TPC and the Time of Flight detectors.\\
After these selections, the signal yield is obtained from a fit to the invariant mass distributions in the different $p_{\rm T}$ bins. 
The 2011 Pb-Pb data sample allowed us to reconstruct the spectra in 10 $p_{\rm T}$ bins in the $p_{\rm T}$ range [1,24] GeV/c for D$^0$, [3,36] GeV/c
 for  D$^+$ and D$^{*+}$, and 3 $p_{\rm T}$ bins in [4,12] GeV/c for D$_s^+$.\\
The  raw yield is corrected for the reconstruction and selection efficiencies, extracted from Monte Carlo
simulations based on $c\bar{c}$ ($b\bar{b}$) pairs generated with PYTHIA with
Perugia-0 tuning and embedded into HIJING Pb-Pb events. \\
A fraction of the total D meson raw yields comes from the decay of B mesons. The contribution of D
mesons coming from B decay was estimated starting from FONLL~\cite{FONLL} predictions
for B cross-sections, which describe well beauty production  at Tevatron~\cite{tev} and LHC~\cite{lhc1,lhc2}. The
$p_{\rm T}$-differential cross section of feed-down D mesons ($\frac{d\sigma^{pp}_{DfromB}}{dp_{\rm T}}$) was
then obtained using the EvtGen \cite{evtgen} package for the B$\rightarrow$D+X decay kinematics.
The non-prompt contribution to the raw yield was finally computed
as:

\begin{center}
\scriptsize
 $\frac{dN}{dp_{\rm T}} = \varDelta_{p_{\rm T}} T_{\rm AA} R_{\rm AA} \frac{d\sigma^{pp}}{dp_{\rm T}}$
\end{center}
where $T_{\rm AA}$ is the nuclear overlap function (from Glauber Model) and the $R_{\rm AA}$ of D mesons from B decays was varied in the range 1/3$<R_{\rm AA}<$3 
since the energy loss of b quarks is not measured.\\
The pp reference (denominator in the $R_{\rm AA}$ formula) was extracted from the 2010 data sample (300 M events) at $\sqrt{s}$ =7 TeV~\cite{7tev}, rescaled 
at 2.76 TeV using the ratio of the FONLL predictions for the D meson cross sections at the two energies. The measured $p_{\rm T}$ differential cross section covers the $p_{\rm T}$ 
range [1,24] GeV/c.
  The results have been validated with the 2011 pp data sample at $\sqrt{s}$ =2.76 TeV~\cite{276}, which covers a narrower $p_{\rm T}$ range ([2,12] GeV/c) due
to low statistics (70 M events).\\
To obtain the p-p reference up to 36 GeV/c an extrapolation based on FONLL/data ratio has been used. \\ 

The upper panel of Fig.1 shows the  $R_{\rm AA}$ of the
4 mesons species as a function of $p_{\rm T}$. For D$^0$, D$^+$ and D$^{*+}$  a suppression of up
to a factor 5 for $p_{\rm T}>$5 GeV/c is observed.

The $R_{\rm AA}$ values of  D$^0$, D$^+$ and D$^{*+}$ have been averaged. This average D meson $R_{\rm AA}$ shows
 shows a similar suppression as that of charged hadrons, as it can be seen in the lower panel of Fig.1. For D$_s^+$, the large statistical and systematic
uncertainties with the present data
sample do not allow to conclude about
low and intermediate $p_{\rm T}$ region, while for 8$<p_{\rm T}<$ 12 GeV/c the measured suppression is similar to the one of non-strange D mesons. Fig. 2 shows
D  meson $R_{\rm AA}$ (data coming from the 2010 data sample~\cite{pbpb}) as a function of centrality compared 
to non-prompt J/$\psi$ (CMS preliminary data) and indicates that c quarks 
are more affected by the medium than b quarks. It should be noted that D meson and J/$\psi$ are measured
in a different  $p_{\rm T}$ and rapidity range.
The comparison of the measured D meson $R_{\rm AA}$ to models, shown
in the bottom panel of Fig. 2, indicates that shadowing alone, NLO(MNR) model, cannot explain such a strong suppression;
models including in-medium parton energy loss can give a reasonable description
of the data.

\begin{figure}
 \centering

   {\includegraphics[width=6cm]{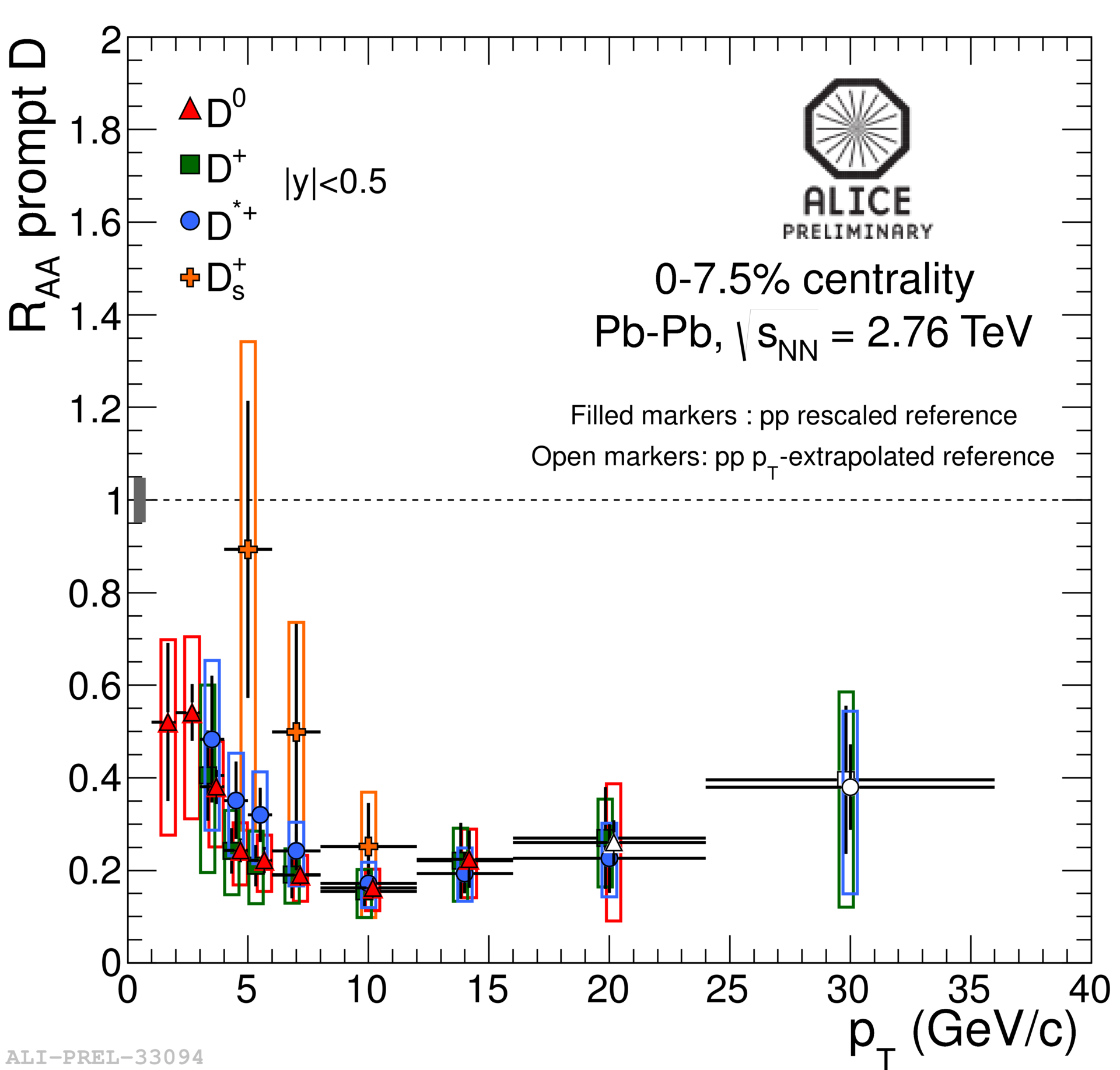}}

   {\includegraphics[width=7cm]{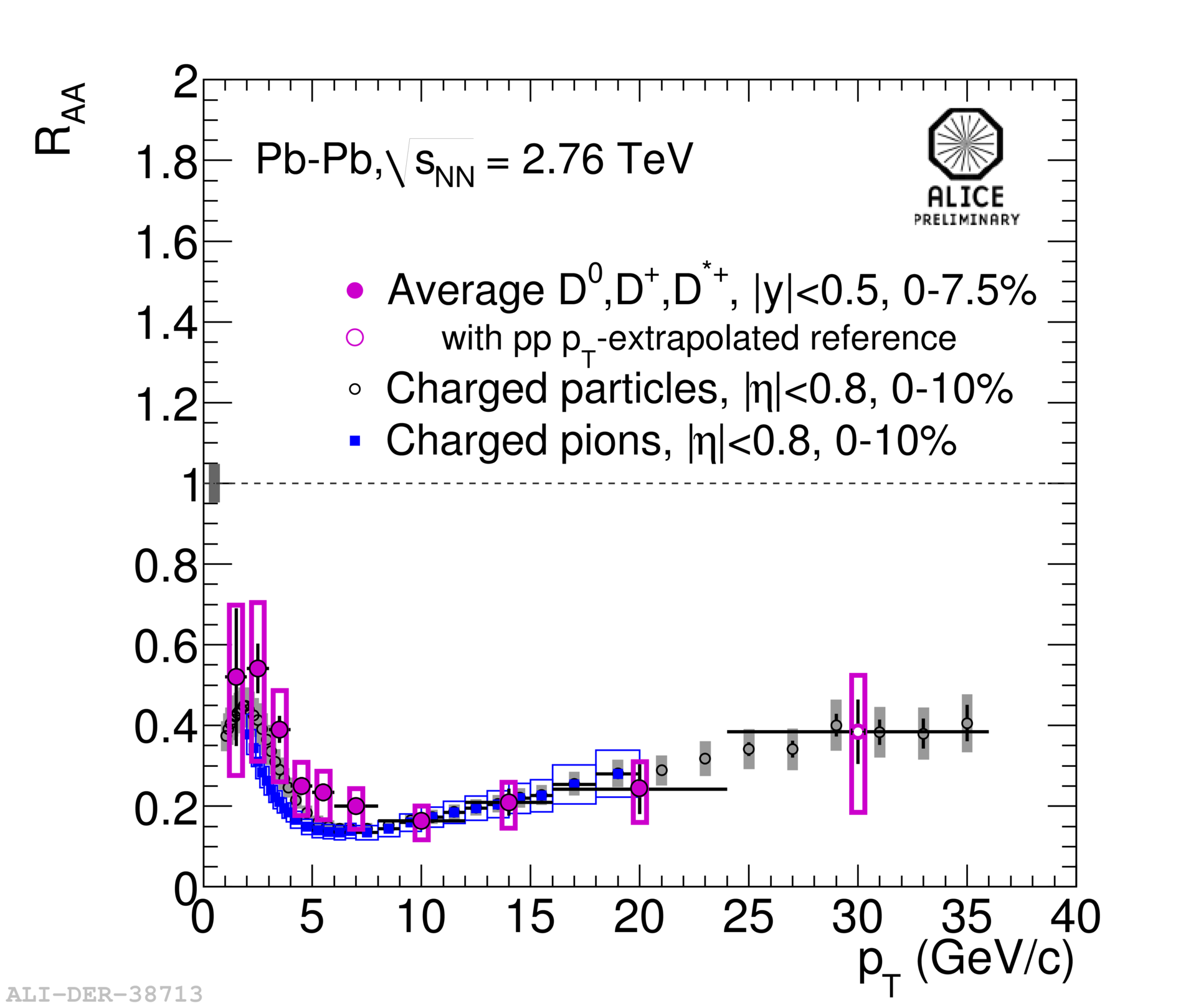}}
 \caption{Upper panel: D$^0$, D$^+$, D$^{*+}$ and D$_s^+$ $R_{\rm AA}$ as a function of $p_{\rm T}$. Lower panel: average D meson $R_{\rm AA}$ as a function of $p_{\rm T}$ compared to charged hadrons $R_{\rm AA}$}
 \end{figure}
\begin{figure}
 \centering

   {\includegraphics[width=7cm]{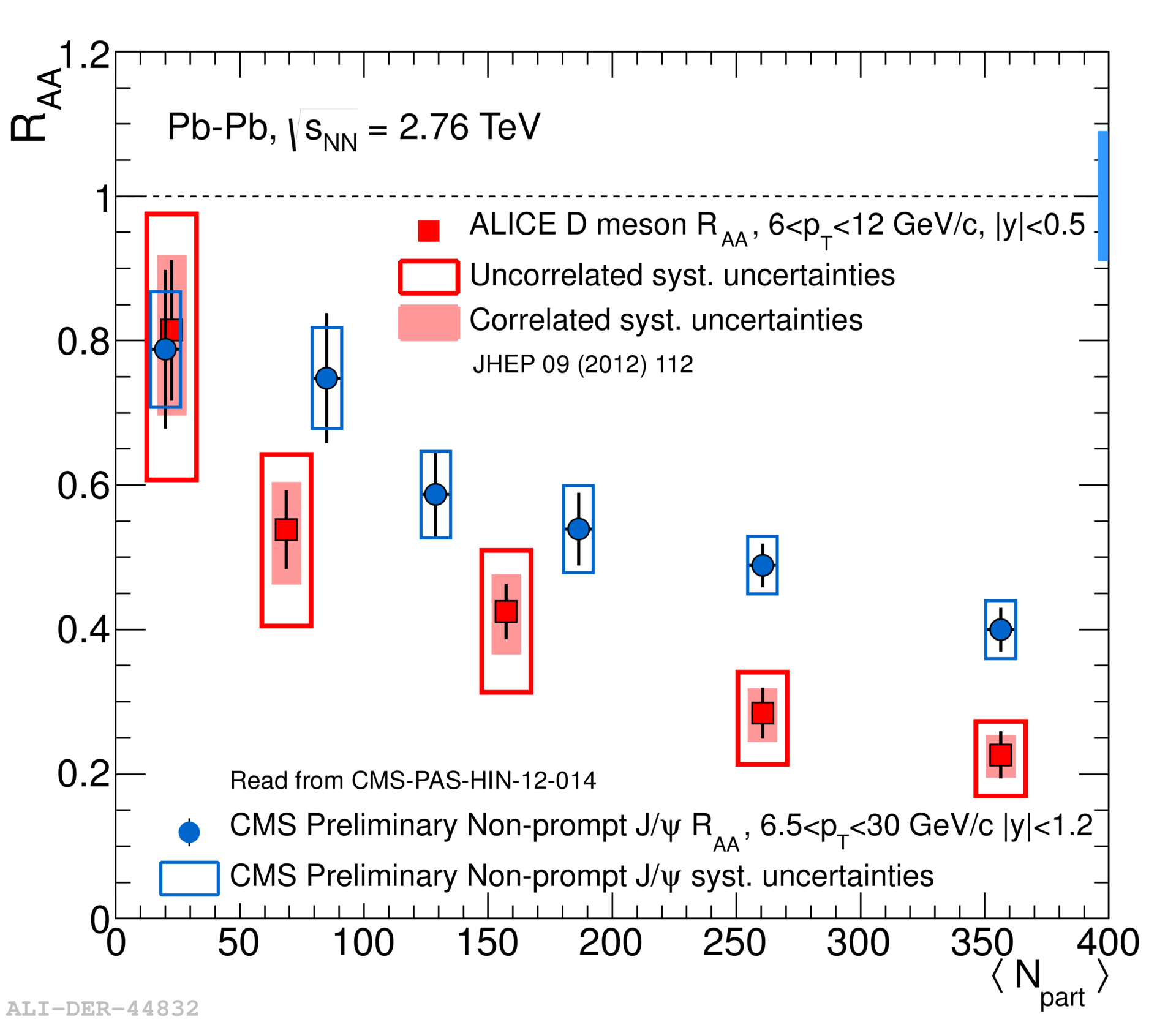}}

   {\includegraphics[width=8cm]{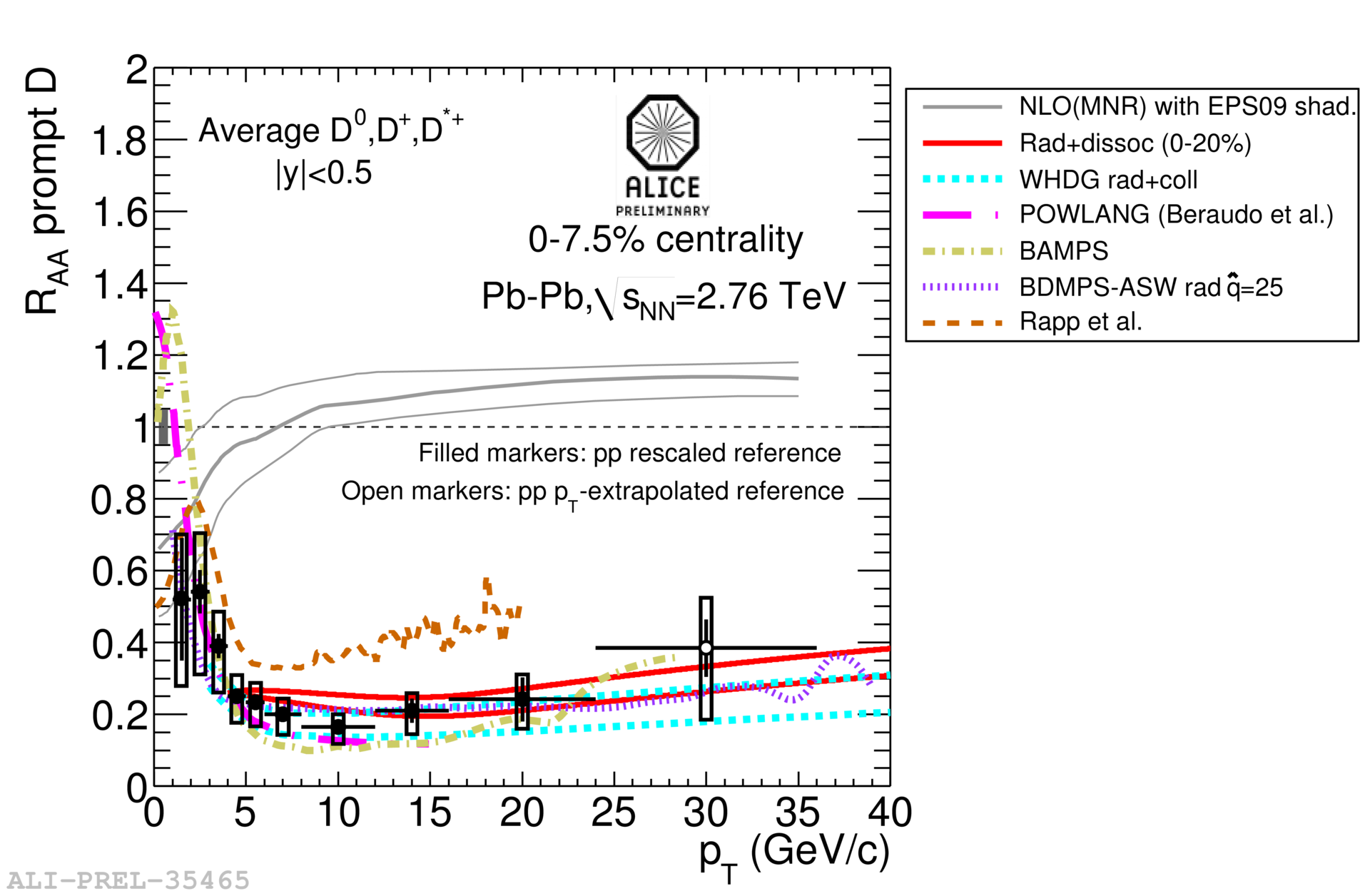}}
 \caption{Upper panel: average D meson $R_{\rm AA}$ as a function of centrality compared to non-prompt J/$\psi$ $R_{\rm AA}$. Lower panel: average D meson $R_{\rm AA}$ as a function of $p_{\rm T}$ compared to models}
 \end{figure}

\subsection{\label{sec:level2} Elliptic Flow}
The v$_2$ analysis was performed in three centrality classes: 0-7.5\%, 15-30\% and 30-50\%. \\
D meson candidates are divided into two sub-samples, depending on the reconstructed D meson azimuthal angle
$\phi$ relative to the event plane $\psi_{\rm EP}$, 
which is an estimator of the reaction plane. $\psi_{EP}$ is computed for each event from the azimuthal distribution of reconstructed TPC tracks.
The two azimuthal regions are:
in plane  for $|\Delta\varphi|<\frac{\pi}{4}$
 and out of plane  for $\frac{\pi}{4}<|\Delta\varphi|<\frac{3\pi}{4}$.
 The invariant mass spectra in the two regions are fitted to get the in-plane and out-of-plane yields, N$_{\rm IN}$ and N$_{\rm OUT}$, 
and the elliptic flow coefficient v$_2$ is calculated as 
\begin{center}
 $v_{2} = \frac{\pi}{4}\frac{N_{\rm IN}-N_{\rm OUT}}{N_{\rm IN}+N_{\rm OUT}}\frac{1}{R_{2}}$
\end{center}
where R$_2$ is the event plane resolution, estimated by reconstructing the event plane angle using two sub-samples of tracks for each event. \\

Results in the centrality range 30-50\%, shown in the upper panel of Fig. 3, indicate non-zero $D^{0}$ v2 (3$\sigma$ for 2$<p_{\rm T}<$6 GeV/c),
 in agreement with that of $D^{+}$ and $D^{*+}$ within uncertaintes.
In addition, $v_2$ is comparable  to that of charged hadrons in the same centrality.\\
An increase of v$_2$ is observed going from central (0-7.5\%) to peripheral (30-50\%) events (central panel of Fig. 3), as expected from the smaller initial geometrical anisotropy in central collisions.\\
The lower
panel of Fig. 3 shows the comparison between the measured D meson $v_2$ as a
function of $p_{\rm T}$ in 30-50\% centrality class, together with  some model predictions. The
measured $v_2$ tends to favour the models that predict larger anisotropy at low
$p_{\rm T}$, suggesting that c quarks participation to the collective motion gives an important contribution to the onset of D meson elliptic flow.

\section{\label{sec:level1} Summary}
From the analysis of the 2011 Pb-Pb data sample collected with the ALICE detector, the reconstruction of D mesons through their hadronic decay channels
has allowed the measurement of D$^0$, D$^+$, D$^{*+}$ and, for the  first time, D$_s^+$ yield in heavy-ion collisions. \\

In order to study the effect
of the medium on the D meson momentum distributions, these results were
compared to a pp reference to compute the nuclear modification factor. The
   R$_{\rm AA}$ of D$^0$, D$^+$ and D$^{*+}$ was
 measured over a wide $p_{\rm T}$ range and shows a suppression up to a factor 5 for  $p_{\rm T}>$5 GeV/c. Conclusions about a possible enhancement
of D$_s^+$ at low-interediate $p_{\rm T}$ still need more statistics, which is expected to be achieved after the LHC and ALICE upgrade plans. \\
D mesons show significantly (3$\sigma$) non-zero $v_{2}$ up to 6 GeV/c for $D^{0}$, $D^{+}$ and $D^{*+}$ in semi-peripheral (30-50\%) collisions with a hint of 
$v_{2}$ decrease going to more central collisions. \\
Different theoretical models can describe  the nuclear modification factor and the elliptic flow separately, while a simultaneous 
 description of both observables is still challenging.

\begin{figure}
 \centering

   {\includegraphics[width=7cm]{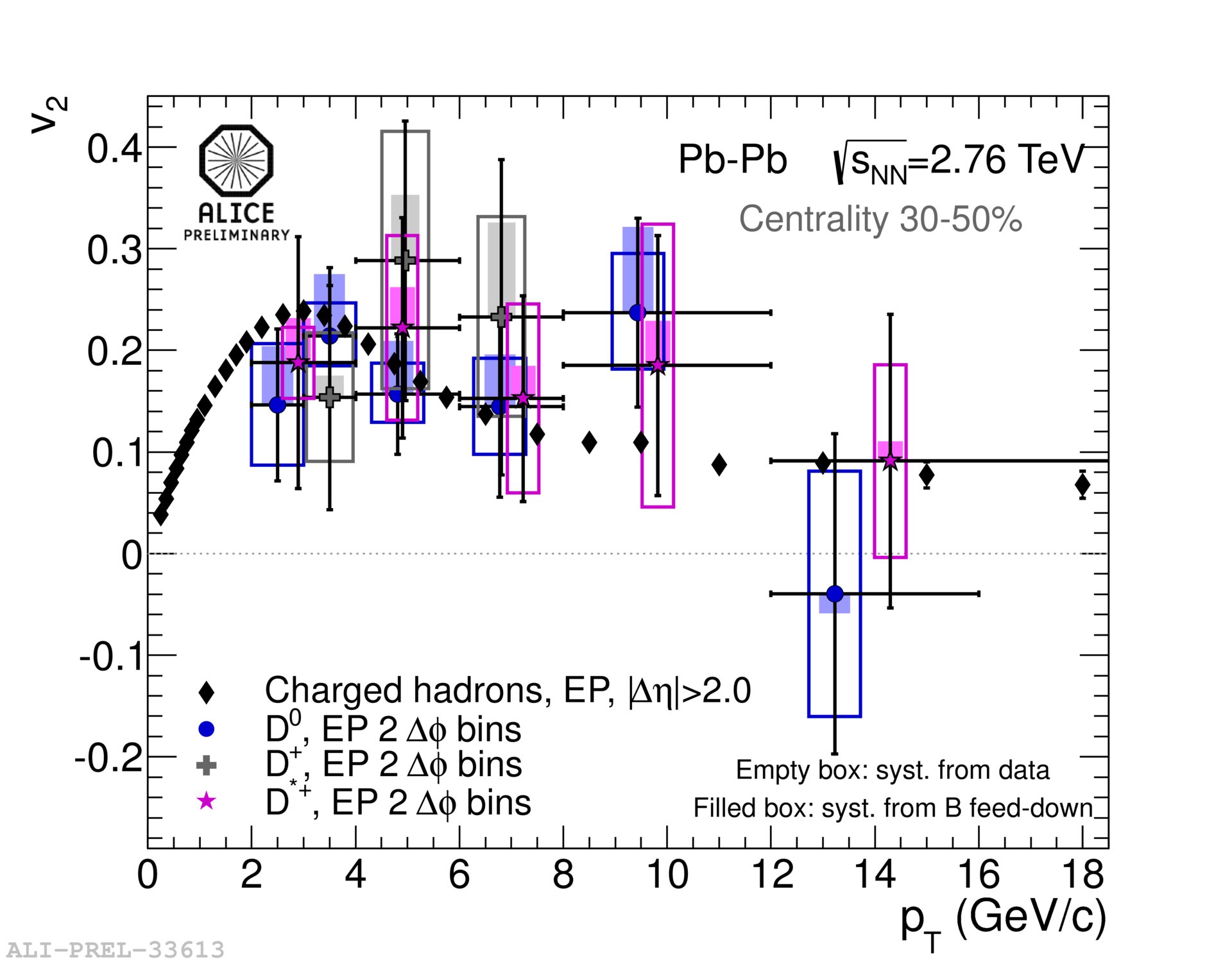}}

   {\includegraphics[width=7cm]{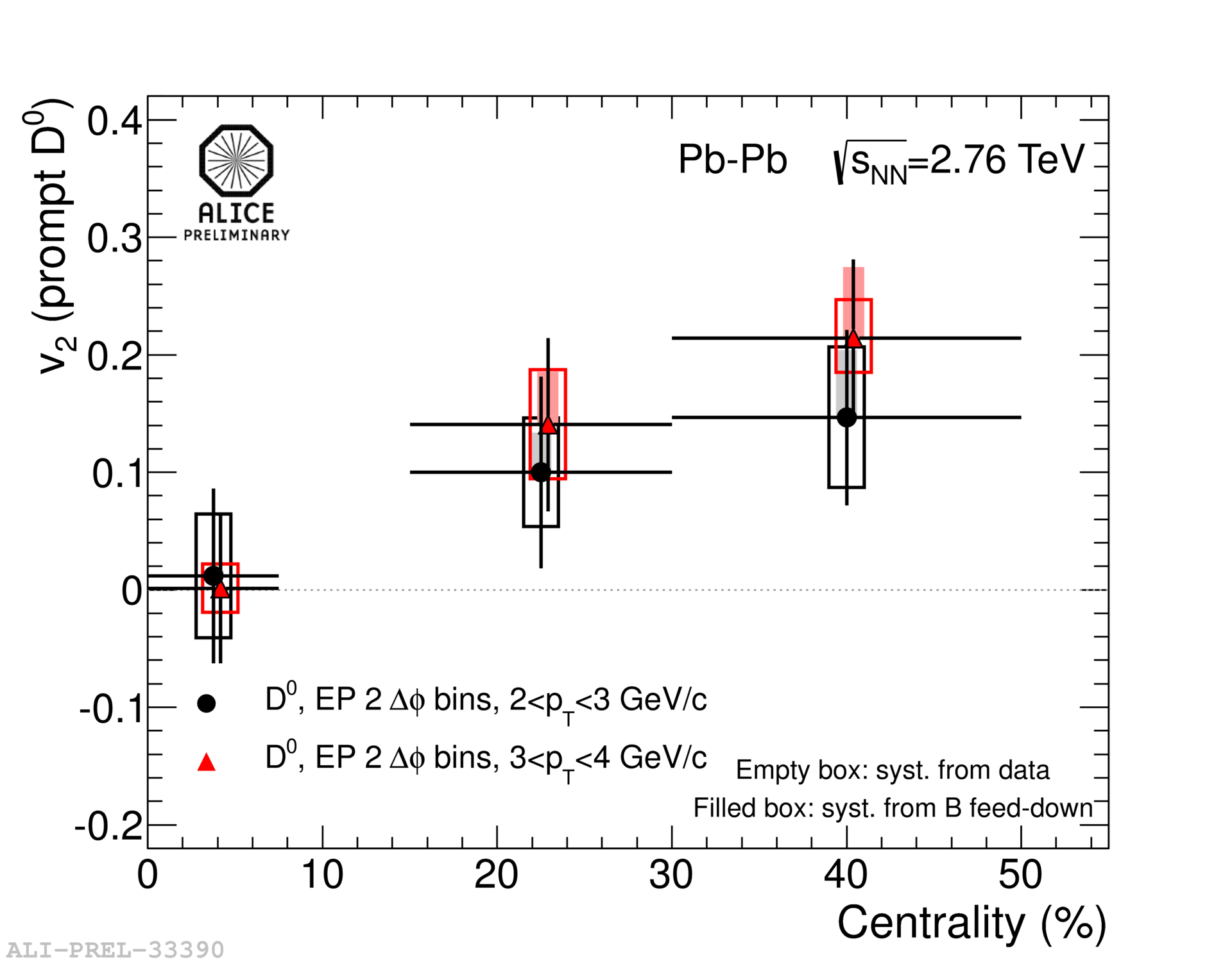}}

   {\includegraphics[width=7cm]{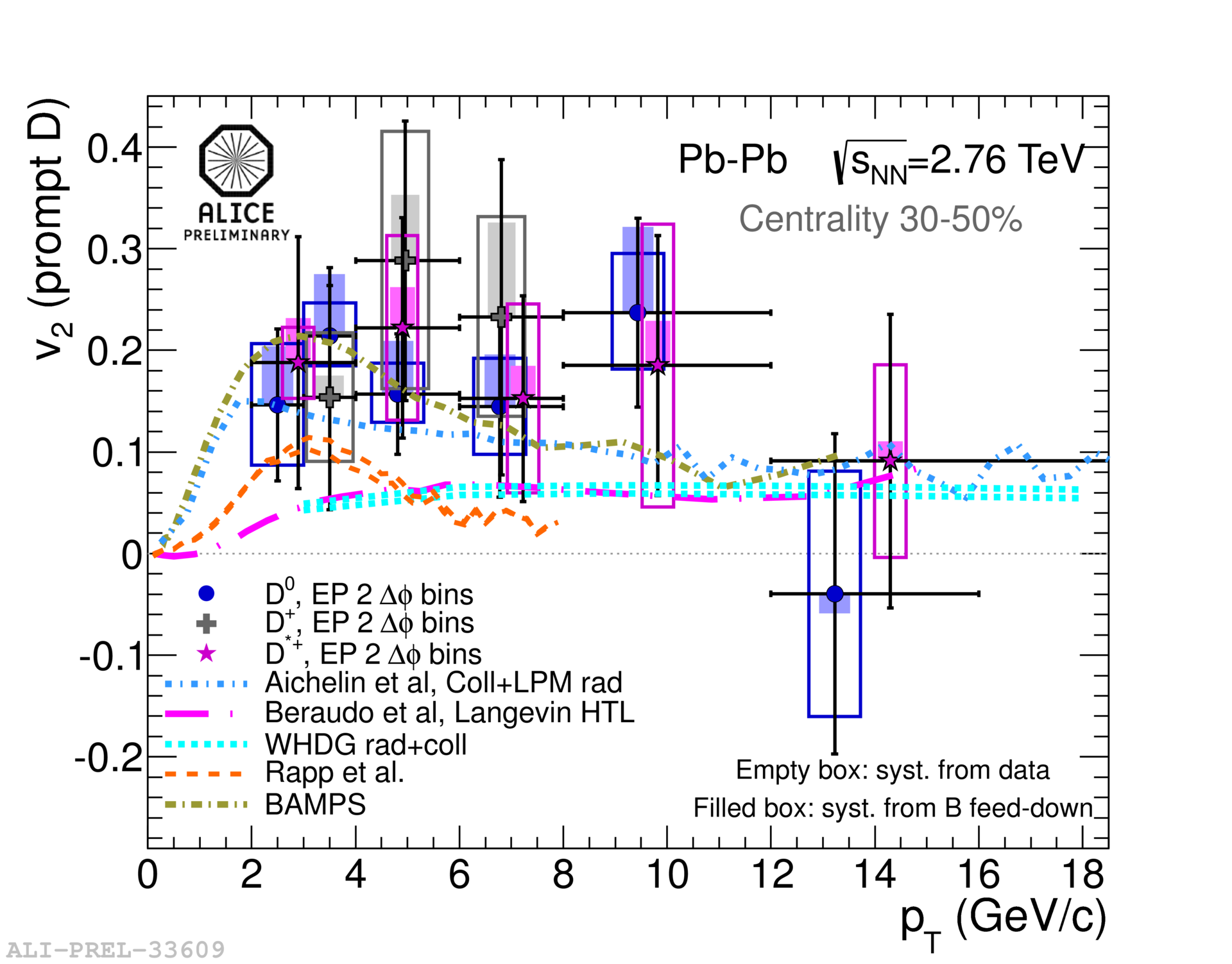}}
 \caption{Upper panel: D$^0$, D$^+$ and D$^*$ v${_2}$ as a function of $p_{\rm T}$ in 30-50\% centrality compared to that of charged hadrons. Middle panel: D$^0$ v${_2}$ as a function of centrality. Lower panel: D$^0$, D$^+$ and D$^*$ v${_2}$ as a function of $p_{\rm T}$ in 30-50\% centrality
compared to models.}
 \end{figure}
\nocite{*}




\end{document}